# Mean stress effect on Gaßner curves interpreted as shifted Wöhler curves

Pietro D'Antuono[a,b,*] Michele Ciavarella[a]

[a]*Department of Mechanics, Mathematics & Management, Polytechnic University of Bari, Via Edoardo Orabona, 4, 70126 Bari BA*

## Abstract

A criterion for the mean stress effect correction in the shift factor approach for variable amplitude life prediction is presented for both smooth and notched specimens. The criterion is applied to the simple idea proposed by the authors in a previous note that Gaßner curves can be interpreted as shifted Wöhler curves. The mean stress correction used has been proposed by Smith, Watson and Topper and, more in general, by Walker. By applying the correction, a new expression for the shift factor G is obtained and, through the application of the theory of the critical distances in its point variant, surprisingly G is demonstrated to be valid for both smooth and notched geometries since it does not seem to depend on the geometry, but only on the fatigue exponent and the loading history. Finally, a comparison with the SAE Keyhole test program data is added to substantiate the findings.

## List of symbols

$a_0$: El Haddad intrinsic defect size.
$a_0(N)$: life dependent crack size.
$a_0^u$: $a_0$ at the ultimate strength.
$A_C$, $B_C$: Ciavarella constants for $a_0(N)$.
$A_{ST}$, $B_{ST}$: Susmel and Taylor constants for $a_0(N)$.
crack/notch characteristic size.
D: damage according to PM rule.
f: geometric factor.
G: shift factor.
k, $C_B$: Basquin's law constants.
$k_{CL}$: slope of the power law in the crack like region for notched specimen.
$K_f$: effective stress concentration factor.
$K_{Ic}$: mode I fracture toughness.
$K_t$: theoretical stress concentration factor.
L/2: critical distance in the TCD-P.
N, S: number of cycles and stress.
$N_B$: number of load blocks in a load history.
$N_e$: cycles to the fatigue limit.
$N_H$: number of cycles in the load history.
$n_j$: number of cycles in a load block j.
$N_j$: fatigue life at load block j.
r: material constant to determine $A_C$, $B_C$.
R: load ratio.
$S'_f$: fatigue strength at one cycle.
$S_{ae}$: effective stress amplitude.
$S_{max}$: maximum stress.
$S_{min}$: mininum stress.
$S_{mj}$, $S_{aj}$: mean and amplitude stress at load block j.
$S_{nom,max}$: maximum nominal alternate stress of the load history.
$S_u$: ultimate tensile strength.
β: multiplicative factor of the spectrum.
ΔK: Stress intensity factor range.
$ΔK_{th}$: Threshold stress intensity factor.
ΔS: stress range.
$ΔS_e$: fatigue limit stress range.
ϰ(N): life dependent stress concentration factor.
$ν_j$: life proportion spent at block j.

## 1 Introduction

The approach to fatigue design [1]–[9] is classically based on the identification of some typical constants, amongst which there is certainly the idea to interpolate with some equation from a value of strength at low number of cycles and another lower strength at high number of cycles, see for instance Weibull [9] or Fuchs and Stephens [4] (become Stephens et al. [3] in a newer edition). Further material constants have been introduced by the advent of Fracture Mechanics by Irwin [10], albeit this generally is treated as a completely different topic from fatigue, although many authors in the decades have tried to find a connection between fatigue and fracture mechanics. For instance Smith and Miller [11] introduced the concept that notches behave like cracks if they are sharp (crack like notches) This concept allows to obtain characteristic diagrams [12]–[16] under the assumption that the governing factor for infinite life modeling is the threshold stress intensity factor range $\Delta K_{th}$, value below which a so-called long crack should not propagate consistently with Paris' law [17]. Indeed, infinite life fatigue design is governed by the threshold stress intensity factor range for long cracks, and by the fatigue limit stress range $\Delta S_e$ for short cracks [18] which implies that to identify the order of magnitude of the critical length of transition from short to long crack, El-Haddad's intrinsic defect size $a_0$ can be introduced

$$a_0 = \frac{1}{\pi}\left(\frac{\Delta K_{th}}{\Delta S_e}\right)^2 \tag{1}$$

*Corresponding author.
*E-mail address*: pietro.dantuono@poliba.it (P. D'Antuono).

The equation is given for a prescribed load ratio $R=S_{min}/S_{max}$. Such transition has been studied in principle experimentally in 1976 by Kitagawa and Takahashi [19] which plotted stress range vs. crack size ($\Delta S$-a) diagrams confirming the validity of the relation connecting the stress intensity factor range $\Delta K$ and the stress range $\Delta S$ that in 1980 was introduced for the first time by El Haddad et al. [20]:

$$\Delta S_e = \Delta K_{th}/\sqrt{\pi(f\,a + a_0)} \tag{2}$$

Where f is a geometric factor. The idea behind Equation (2) is one aspect of the "Theory of the Critical Distances" whose ancestor, as summarized by Taylor [21], and Yao et al. [22], can be identified in the effective stress concentration factor $K_f$ proposed by Neuber [23], Kuhn and Hardrath [24] who in the early '50s assumed that the notched specimen fails if the averaged stress over the distance $A_{KH}$ ahead of the notch root is equal to the fatigue limit $S_e$ of the plain specimen. Finite life design is a more complicated matter and can be approached in multiple ways: (i) Paris' law could be directly integrated with opportune correction for short crack effect, as evidenced by Pugno et al [25] and by Ciavarella and Monno [26]or (ii) a finite life Kitagawa-Takahashi diagram with short crack effect could be introduced, as suggested by Ciavarella [27] or by Maierhofer et al. [28]. In particular, Ciavarella [27] attempted a generalized finite life form of the El Haddad equation, postulating a life dependent power law for the intrinsic defect size, i.e. $a_0=a_0(N)$ (life-dependent)

$$a_0(N) = \frac{1}{\pi}\left(\frac{\Delta K_{th}}{\Delta S_e}\right)^2 (N/N_e)^{2\left(\frac{1}{k}-\frac{1}{r}\right)} = A_C \cdot N^{B_C} \tag{3}$$

Where $B_C=2(1/k-1/r)$, $A_C=1/\pi\cdot(\Delta K_{th}/\Delta S_e)^2\cdot N_e^{-B_C}$, $N_e$ is the number of cycles to the fatigue limit, k is Basquin's law exponent and r is a material constant which could coincide with the Paris' law exponent since it has been introduced to the postulate the evolution of $\Delta K_{th}(N)= \Delta K_{th}\cdot(N_e/N)^{1/r}$. A very similar approach to the problem has been proposed by Susmel and Taylor [21], [29]–[32] in their theory of the critical distances in its point variant (TCD-P) in which a life dependent power law evolution of the critical distance L/2 from the crack tip/notch root has been postulated, i.e.

$$\frac{L(N)}{2} = \frac{a_0(N)}{2} = A_{ST} \cdot N^{B_{ST}} \qquad (4)$$

Where the constants $A_{ST}$ and $B_{ST}$ have similar meaning to $A_C$ and $B_C$ and can be determined both through experimental fitting of data from notched specimens and from basic material properties. Considering the derivation from the material properties, it is straightforward to define the power law from the El Haddad intrinsic defect size to the equivalent ultimate, or static, one

$$a_0^u = 1/\pi \left(\frac{K_{Ic}}{S_f'}\right)^2 \qquad (5)$$

Where $K_{Ic}$ is the mode I fracture toughness and $S'_f$ is the fatigue strength at one cycle according to Basquin's law, i.e. $S^k N = C_B$. $S'_f$ has been used instead of the ultimate tensile strength $S_u$ since it is there's still another free parameter to be set: the number of cycles $N_u$ where the $a_0^u$ is assumed to hold (see Ciavarella et al. [33] for further explanations).

## 1.1 Variable amplitude fatigue

Fatigue under cyclic loading with a constant amplitude and a constant mean load is addressed as constant-amplitude (CA) fatigue loading, a classical example of which is the sinusoidal loading applied in many fatigue tests. Nevertheless, many components undergo complex load histories in their operating life, called variable-amplitude (VA) loading. The study of this phenomenon is still of great interest both in academia and in industry, and this is confirmed by the fact that multiple authors keep studying this topic although there have been thousands of experimental campaigns, analytical and numerical models that tried to predict the VA fatigue behavior of materials in the last century. The simplest VA fatigue prediction rule has been proposed for the first time in 1924 by Palmgren [34] for the fatigue calculation of ball-bearings. Supposing that the load history is made of $N_B$ load blocks, each one containing $n_j$ cycles at the stress amplitude $S_{mj}$, $S_{aj}$ and the corresponding fatigue life $N(S_{aj})=N_j$, the rule can be expressed as

$$D = \sum n_j/N_j = 1 \qquad (6)$$

In other words, Palmgren postulated the linear accumulation of the fatigue damage, stating that the failure occurs when the damage D=1. Anyway Palmgren did not provide a derivation for the rule, and the same holds for Langer [35] that in 1937 postulated the same rule applied separately to the crack initiation and to the crack propagation phases. The first derivation of the linear damage accumulation rule has been proposed by Miner [36]. His hypothesis was that the work that can be adsorbed until failure is a constant value and that the amount of work adsorbed during $n_j$ is directly proportional to $n_j$. Thus, said W the total work and $w_j$ the work adsorbed during the block $n_j$, the criterion is $\Sigma_j w_j = W$. The use of Miner hypothesis ($n_j/N_j = w_j/W$) leads immediately to Equation (6). Miner conducted a series of tests on smooth and riveted 2024-T3 aluminum alloy sheet specimens by applying load histories having $2 \leq N_B \leq 4$ and found $0.61 \leq \Sigma_j n_j/N_j \leq 1.45$, very close to 1 on average. Since then the linear damage accumulation rule has been addressed very often as Miner's rule, but probably Palmgren-Miner's (PM) rule, is the more corrected form and it is how the rule will be called in this work. Since that time many works have been published to verify the PM rule and to find its limits of validity. For example, also Ciavarella et al. [37] have shown that the limit values of PM rule range from 0.001 to 10. Since then several theories that tried to overcome this limit and generalize the rule have been proposed. Some of them where quite simple, e.g. Leve [38] in 1960 postulated the first simple nonlinear damage accumulation rule $\Sigma_j (n_j/N_j)^{c_L}$ with $c_L > 1$, whilst others were much more complicated, like Park and Padgett's [39] general class of cumulative damage models which defined the damage as function of a statistical "strength reduction function". Despite all the interest in defining a generalized damage accumulation model, PM remains by far the most used rule in fatigue design. Thus, in order to increase its conservativity some handbooks, e.g. the FKM-Guideline [40], suggest reducing the critical damage from 1 to 0.3 for steels, steel castings, aluminum alloys, while keeping 1 for ductile iron, gray cast iron, malleable cast iron, albeit Sonsino [41]–[43] and Schijve [44], [45] suggest that testing is always the best choice. However, even testing can be (i) extremely expensive and (ii) very difficult both in the setup of the VA experiments and in the interpretation of the results. Indeed, carrying a VA fatigue test campaign is an art on its own involving the concepts of safety factors both in life and in stress [46]: essentially, under a given service loading history, a single test can assess a given same reliability (typically assuming a Weibull distribution) only by increasing the load or the number of cycles/blocks to failure with respect to the mission. The former version is preferred because of the obvious time (and cost) savings that it implies, although special attention is needed when testing at high loads; indeed, during the test of a complex component with a complex loading test rig local plasticization phenomena (not foreseen via previous finite elements analyses) might arise. On the other hand, a VA test with higher expected cycles to failure might (and usually it does) last too long if the bandwidth of the rig actuators is limited with respect to the load amplitude they should provide, plus if time-dependent phenomena affect the fatigue process results obtained with a too high loading frequency would be partially or completely unreliable. Often small amplitude cycles are omitted for simplicity and to accelerate testing (and similarly in the original PM rule the cycles below fatigue limit are omitted from the computation of the damage), although in some design handbooks, especially in welded joints, these are known to produce fatigue damage. Indeed, low amplitude cycles can be dealt with according to the PM rule with prolongation of the Wöhler curve below the knee point with the same slope, or according to Haibach [47] with a reduced slope. The former method is the most conservative amongst the listed ones and has been adopted in the current methodology without loss of generality, i.e. for VA fatigue calculations the S/N power law curve for CA is supposed to extend for N→∞. This is particularly true for materials which do not show a clear fatigue limit, like aluminum or especially magnesium alloys, for which Haibach correction would not be required anyway. At the other extreme, it is demonstrated that the application of some isolated high amplitude cycles has a beneficial effect on the total life since it induces compressive residual stresses at notch roots. These effects are not considered in the current methodology for sake of conservativity; furthermore, the model is suitable for fast and simple design level assessment, hence it is in the authors' intent to keep it as slender as possible. Moreover, in many cases, it is still debatable whether load spectra are known with satisfying accuracy, or if cycle-counting methods (such as rainflow or range-pair) are reliable (i.e. if load sequence effects are not important); therefore Miner's law is still very much used, and this is why predictions cannot completely substitute testing. They are just going to suggest better ways to plot Gaßner curves than what presently done, or and what one may expect when applying PM rule according to the theory of the critical distances (TCD) in complex situations, perhaps coming from finite element results of the stress fields, though in the case studies analyzed finite elements have been avoided through analytical considerations.

## 2 Shift factor definitions

### 2.1 Gaßner curves for smooth specimen

The definition of shift factor has already been given in a previous paper by Ciavarella et al. [48]. It is based on the hypothesis that Basquin's law holds

$$N_u S_u^k = N_e S_e^k = N S^k = C_B \tag{7}$$

Where the equation has been written also at the extreme points of the classical domain of validity of Basquin's law, i.e. at some low number of cycles $N_u$ corresponding to the static strength $S_u$ and at a high number of cycles $N_e$ corresponding to the fatigue limit $S_e$. In the VA case the existence of the fatigue limit is disregarded and the Wöhler curve is extended to infinity. According to PM rule, the damage D in function of the alternate stress $S_a$ for a given load history containing $N_B$ blocks and a total number of cycles $N_H$ would be

$$D = \sum_{j=1}^{N_B} \frac{n_j}{N_j} = \sum_{j=1}^{N_B} \frac{n_j}{N_H} \cdot \frac{N_H}{N_j} = N_H \sum_{j=1}^{N_B} \frac{\nu_j}{N_j} = \frac{N_H}{C_B} \sum_{j=1}^{N_B} \nu_j S_{aj}^k \tag{8}$$

Where $\nu_j = n_j/N_H$ is the proportion of cycles spent at level j on the total number $N_H$. The life under the sum of all the $N_B$ blocks is $\overline{N}$

$$\frac{1}{\overline{N}} = \frac{D}{N_H} = \frac{1}{C_B} \sum_{j=1}^{N_B} \nu_j S_{aj}^k \tag{9}$$

Therefore, normalizing the history by its peak tension $S_{a.max}$ such that $\overline{S} = \beta S_{a,max}$ (and $\overline{S}_{aj} = \beta S_{aj}$) and by varying the factor $\beta$ a full Gaßner curve is obtained

$$\frac{1}{\overline{N}(\overline{S})} = \frac{\overline{S}^k}{C_B} \sum_{j=1}^{N_B} \nu_j \left(\frac{S_{aj}}{S_{a,max}}\right)^k = \frac{\overline{S}^k}{C_B} \cdot G \qquad (10)$$

Where G has been addressed as shift factor and has the following expression

$$G = \sum_{j=1}^{N_B} \nu_j \left(\frac{S_{aj}}{S_{a,max}}\right)^k \qquad (11)$$

G depends on the spectrum and the fatigue exponent only. In this way the Gaßner curve for a smooth specimen can be interpreted as a shifted Wöhler curve in the Log(S)/Log(N) coordinates. Equation (10) can be rewritten as

$$\left(\frac{\overline{S}}{G^{-\frac{1}{k}}}\right)^k \overline{N} = C_B \qquad (12)$$

Besides, the Gaßner curve can be plotted as overlapped to the Wöhler curve by using the scale $\mathrm{Log}(S/G^{-1/k})$ instead of the common $\mathrm{Log}(S)$.

## 2.2 Gaßner curve in the crack like region

In a previous work Ciavarella et al. [33] showed that in many cases when notches are sufficiently "sharp" their behavior is not too dissimilar from cracks in a certain span of fatigue cycles ranging from the quantities $N_0$ and $N^*$ depending on the TCD-P constants, viz.

$$N_0 = \left(\frac{a}{A_{ST}}\right)^{\frac{1}{B_{ST}}} \qquad (13)$$

$$N^* = \left(\frac{a}{K_t^2 A_{ST}}\right)^{\frac{1}{B_{ST}}} = \frac{N_0}{K_t^{2/B_{ST}}} = N_e \left(\frac{a}{K_t^2 a_0}\right)^{\frac{1}{B_{ST}}} = N_e \left(\frac{a}{a^*}\right)^{\frac{1}{B_{ST}}} \qquad (14)$$

Where $K_t$ is the theoretical stress concentration factor. Then this hypothesis has been generalized to a wider family of problems through Susmel and Taylor's [21], [29]–[32] TCD-P. The aim of Ciavarella et al. [33] was the formulation of an analytical S/N curve model which could account for the effect of notches in medias to estimate their fatigue life under CA loading with satisfying accuracy. In the current work, under the proper hypotheses, such model will be extended to the estimation of VA fatigue life by demonstrating that the shift factor G is not affected neither by the presence of notches nor by non-null mean stresses, thence it can be applied also to the more sophisticated S/N curves previously defined. For instance, the asymptotic part ($x \rightarrow 0$) of the Westergaard [49] solution for a crack of length 2a immersed in an infinite plate and subjected to opening mode loading with asymptotic nominal stress $S_{nom}$ is

$$S(x) = \frac{K_I}{\sqrt{2\pi x}} = \frac{S_{nom}}{\sqrt{2 x/a}} \tag{15}$$

The TCD-P suggests evaluating the stress at a distance $a_0(N)/2$ from the crack, resulting in $S(x(N))$ to evaluate the fatigue life. This means that there is a spectrum of $S_{aj}$ values giving a spectrum of $S(x(N))$ values, where one takes (either Ciavarella [27] or Susmel and Taylor variants [21], [29]–[32]) a critical distance of the form

$$x(N) = \frac{a_0(N)}{2} = A_{C-ST}\, N^{B_{C-ST}} \tag{16}$$

Where the subscript C-ST stands for Ciavarella-Susmel and Taylor (anyway ST is going to be used here, congruently with Ciavarella et al. [33]). Equation (14) constants shall be calibrated either according to some dedicated tests or with some material constants. Substituting (16) into (13) gives

$$S_j(x(N)) = \frac{S_{nom,j}}{\sqrt{\frac{A_{ST}}{a} N^{B_{ST}}}} \tag{17}$$

In order to use Equation (17) in VA loading, it must be hypothesized that the intrinsic defect size $a_0$ is not dependent on the spectrum, but only on the final life of the specimen for a given load history. Hence, using PM rule

$$\frac{1}{\overline{N}} = \frac{D}{N_H} = \frac{1}{C_B} \sum_{j=1}^{N_B} v_j\, S_{nom,j}^k \left(\frac{A_{ST}}{a} \overline{N}^{B_{ST}}\right)^{-\frac{k}{2}} = \frac{1}{C_B} \left(\frac{A_{ST}}{a}\right)^{-\frac{k}{2}} \overline{N}^{-\frac{B_{ST}k}{2}} \sum_{j=1}^{N_B} v_j\, S_{nom,j}^k \tag{18}$$

Notice that this is explicit in $\overline{N}$

$$\overline{N}^{\frac{B_{ST}k}{2}-1} = \frac{1}{C_B} \left(\frac{A_{ST}}{a}\right)^{-\frac{k}{2}} \sum_{j=1}^{N_B} v_j\, S_{nom,j}^k \tag{19}$$

Therefore, normalizing the history by the peak tension $S_{nom,max}$ so that $\overline{S}=\beta S_{nom,max}$ the Gassner curve in the crack like region is

$$\left(\overline{N}(\overline{S}_{nom})\right)^{\frac{B_{ST}k}{2}-1} = \frac{\overline{S}_{nom}^k}{C_B}\left(\frac{A_{ST}}{a}\right)^{-\frac{k}{2}} \sum_{j=1}^{N_B} v_j \left(\frac{S_{nom,j}}{S_{nom,max}}\right)^k = \frac{\overline{S}_{nom}^k}{C_B}\left(\frac{A_{ST}}{a}\right)^{-\frac{k}{2}} G \qquad (20)$$

Where the shift factor G is the same as for the smooth material. However, notice that the new curve can be written as

$$\overline{S}_{nom}^{k_{CL}} \; \overline{N} = \left(\left(\frac{A_{ST}}{a}\right)^{-\frac{k}{2}} \cdot \frac{G}{C_B}\right)^{-k_{CL}/k} \qquad (21)$$

Where the new slope $k_{CL}=k/(1-B_{ST}k/2)$. Consequently, this curve would have the same slope as the smooth one only if $B_{ST}=0$, i.e. if $a_0$ stayed constant. As done for Equation (11), Equation (21) can be rearranged as

$$\left(\frac{\overline{S}_{nom}}{G^{-\frac{1}{k}}}\right)^{k_{CL}} \overline{N} = \left(\left(\frac{A_{ST}}{a}\right)^{-\frac{k}{2}} \cdot \frac{1}{C_B}\right)^{-k_{CL}/k} \qquad (22)$$

From Equation (22), a bigger crack like notch implies a shorter life, consistently with what expected. Besides, this equation means that, in terms of nominal stress amplitude the S/N curve in the crack like region is shifted exactly of the same amount of the unnotched S/N curve when obtaining VA data.

### 2.3 Gaßner curve in the blunt notch region

Considering the piecewise power law defined by Ciavarella et al. [33] for Wöhler curves, there is a region for $N>N^*$ where the fatigue behavior can be approximated with enough accuracy by the Wöhler curve of the smooth specimen reduced by the stress concentration factor, i.e.

$$(S\,K_t)^k\,N = C_B \qquad (23)$$

Or

$$S^k\,N = C_B\,K_t^{-k} \qquad (24)$$

This region is addressed as blunt notch region, as also stated by Ciavarella [50]. Through the same passages made here above, the Gaßner curve in the blunt notch region becomes

$$\left(\frac{\overline{S}}{G^{-\frac{1}{k}}}\,K_t\right)^k \overline{N} = C_B \qquad (25)$$

Being G the same shift factor defined for the smooth material. This means that in the entire domain of N, the Gaßner curve for a notched body defined through the TCD-P (and approximated with a piecewise power law) is equivalent to a Wöhler curve shifted by $G^{-1/k}$. It is noteworthy that in this work $K_t$ and $K_f$ are used almost equivalently because in the next examples the notch radius is way larger than $a_0(N)$, therefore $K_f \rightarrow K_t$.

## 2.4 Gaßner curve for the general equation of a notch

From the definition of the exact TCD-P Wöhler curves for a crack and a circular notch the following general S/N equation is retrieved

$$N\left(\bar{S}\,\kappa(\bar{N})\right)^k = C_B \tag{26}$$

Being $\kappa(N)$ a life dependent stress concentration factor. Through the application of the same procedure that has been shown for the crack like and blunt notch regions, a general definition of Gaßner curve for an S/N curve with a life dependent stress concentration factor can be written

$$N\left(\frac{\bar{S}}{G^{-\frac{1}{k}}}\kappa(\bar{N})\right)^k = C_B \tag{27}$$

Where $\kappa(\bar{N})$ can be (25)

$$\begin{aligned}
\kappa_W(N) &= \frac{1 + \frac{a_0(N)/2}{a}}{\sqrt{\frac{a_0(N)/2}{a}}\sqrt{2 + \frac{a_0(N)/2}{a}}} & \text{crack} \\
\kappa_K(N) &= \frac{1}{2}\cdot\left(2 + \left(\frac{a}{a + \frac{a_0(N)}{2}}\right)^2 + 3\left(\frac{a}{a + \frac{a_0(N)}{2}}\right)^4\right) & \text{hole}
\end{aligned} \tag{28}$$

Therefore the iterative procedure suggested by Susmel and Taylor [32] for VA life computing seems not needed.

## 3 Mean stress effect on the shift factor

### 3.1 A brief overview on mean stress effect corrections

The study of mean stress effect on fatigue loading is probably almost as old as the study of fatigue itself. Indeed, in the very old days of fatigue, Wöhler [51]–[53] between 1858 and 1870 already mentioned a possible detrimental effect of positive mean cyclic stresses on life of railway axles. The first quantitative model relating the stress amplitude $S_a$ and mean stress $S_m$ through the ultimate tensile strength $S_u$ dates 1874 from Gerber [54], who introduced the famous Gerber parabola

$$\left(\frac{S_m}{S_u}\right)^2 + \frac{S_a}{S_{ae}} = 1 \tag{29}$$

Where $S_{ae}$ is the effective stress amplitude in fully reversed loading conditions. In this model, the equivalent stress amplitude can be expressed as a function of the mean and alternate stress, being the ultimate tensile strength a material constant. This means that for a given couple $(S_m, S_a)$ there exists an equivalent condition $(0, S_{ae})$ (fully reversed loading) which provides the same number of cycles to failure as $(S_m, S_a)$. Forty years later, in 1914, Equation (29) has been "replaced" by the modified Goodman [55] line

$$\frac{S_m}{S_u} + \frac{S_a}{S_{ae}} = 1 \tag{30}$$

Which today is (maybe because of its simplicity) the most commonly used mean stress correction in industry and probably the most "popular" model with engineering students. In 1939 a more conservative version of the Goodman line has been proposed by Söderberg [56] who replaced the ultimate tensile strength with the yielding stress of the material, i.e.

$$\frac{S_m}{S_y} + \frac{S_a}{S_{ae}} = 1 \tag{31}$$

Anyway, the Söderberg correction is considered by many authors way overconservative, in fact Woodward et al. [57] stated *"The Söderberg line is safe for nearly all materials, but in very many instances the line seriously over-estimates the effect of mean stress"*. Moreover, it has been demonstrated that for many types of steels even the modified Goodman line (and consequently Söderberg line) provides too conservative corrections [58], [59], and for this reason sometimes it is replaced by the Morrow [60] line which substitutes the ultimate tensile strength with the fatigue strength at one cycle, namely

$$\frac{S_m}{S'_f} + \frac{S_a}{S_{ae}} = 1 \tag{32}$$

$S'_f$ is not much higher than $S_u$ for materials that do not exhibit a pronounced necking, thus Goodman and Morrow lines provide similar corrections. However, in the case of materials which show high plastic deformations, the fatigue strength at one cycle can be much higher than the ultimate tensile strength resulting in a highly less conservative Morrow line with respect to Goodman's. Dowling [58] in his Figure 3 and Figure 4 has shown this phenomenon, with Morrow correction giving highly non-conservative estimates in the case of 2024-T3 aluminum, while providing very good estimates in the case of AISI 4340 steel. It is very common to find Equations (29)–(32) expressed as a function of the effective stress amplitude $S_{ae}$ which in fact usually is the unknown of the problem. In 1970 the Smith-Watson-Topper [61] (SWT) proposed an equation where the mean stress effect was not dependent on any material properties, but only on the loading history itself. The model can be written equivalently as:

$$\begin{aligned} S_{ae} &= \sqrt{S_{max} S_a} &\text{(a)} \\ S_{ae} &= S_{max}\sqrt{\frac{1-R}{2}} &\text{(b)} \\ S_{ae} &= S_a\sqrt{\frac{2}{1-R}} &\text{(c)} \end{aligned} \qquad (33)$$

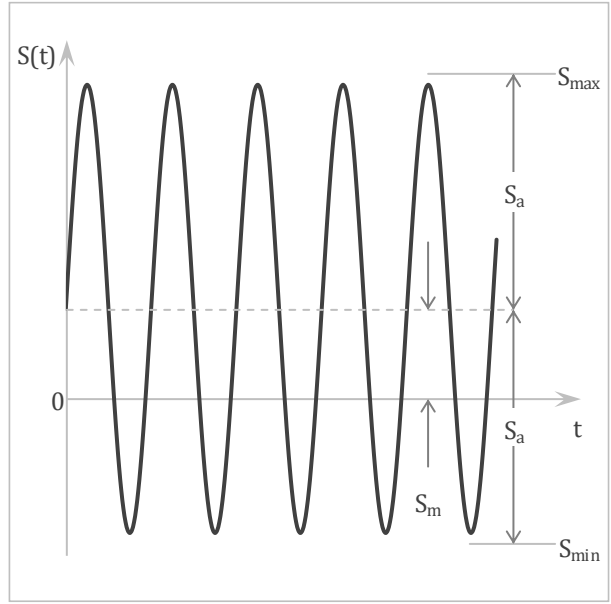

Figure 1 – Definitions for fatigue stress cycle

Where $S_{max}$ is the maximum stress in the cycle and R is the load ratio. Equations (33) (a), (b) and (c) are equivalent since $S_a=½·S_{max}·(1-R)$, as can be easily recovered from Figure 1, where an example constant amplitude cyclic loading history has been plotted. A generalization of SWT model is the one from Walker [62] which can be interpreted as a modified SWT model with a fitting exponent γ. Therefore, Walker equation can be written similarly to Equation (33), i.e

$$\begin{aligned} S_{ae} &= S_{max}^{1-\gamma} S_a^{\gamma} &\text{(a)} \\ S_{ae} &= S_{max}\left(\frac{1-R}{2}\right)^{\gamma} &\text{(b)} \\ S_{ae} &= S_a \left(\frac{2}{1-R}\right)^{1-\gamma} &\text{(c)} \end{aligned} \qquad (34)$$

Which for γ=½ obviously returns Equation (33). Dowling [58] in his Figure 5 and Figure 6 shows that for Al 2024-T3 and AISI 4340 steel the Walker and SWT equations overcome the limitations of Goodman and Morrow lines discussed above related with the ductility of the material. Dowling concludes his work stating that Walker and SWT models are the most accurate for general use, precising obviously that Walker model gives higher accuracy when the exponent γ is known or can be estimated. In the current work SWT and Walker models are going to be used since, besides Dowling conclusions, they are independent of the material constants, which means that they can lead to more general conclusions on the shift factor definition for variable amplitude life prediction.

## 3.2 Generalized shift factor with mean stress effect

The definition of G given in § 2 does not account for the crucial mean stress effect, therefore it is only suitable for variable amplitude fatigue in fully reverse loading conditions. In order to introduce a mean stress effect correction in the definition of the G, Equation (8) shall be rewritten in terms of effective stress amplitude, viz.

$$D = \sum_{j=1}^{N_B} \frac{n_j}{N_j} = \sum_{j=1}^{N_B} \frac{n_j}{N_H} \cdot \frac{N_H}{N_j} = N_H \sum_{j=1}^{N_B} \frac{\nu_j}{N_j} = \frac{N_H}{C_B} \sum_{j=1}^{N_B} \nu_j \, S_{aej}^k \tag{35}$$

### 3.2.1 Smith Watson Topper mean stress effect correction

SWT mean stress correction (Equation (32)) shall be substituted into the PM rule (Equation (35)). With this correction Equation (35) becomes

$$\frac{1}{\overline{N}} = \frac{D}{N_H} = \frac{1}{C_B} \sum_{j=1}^{N_B} \nu_j \left( S_{maxj} \, S_{aj} \right)^{k/2} \tag{36}$$

And normalizing the history by its peak tension $S_{a.max}$ such that $\overline{S} = \beta S_{a,max}$ (and $\overline{S}_{aj} = \beta S_{aj}$, $\overline{S}_{maxj} = \beta S_{maxj}$) and by varying the factor β the full Gaßner curve is obtained again as

$$\frac{1}{\overline{N}(\overline{S})} = \frac{\overline{S}^k}{C_B} \sum_{j=1}^{N_B} \nu_j \left( \frac{S_{maxj} \, S_{aj}}{S_{a,max}^2} \right)^{k/2} = \frac{\overline{S}^k}{C_B} \cdot G \tag{37}$$

Where G has a slightly different definition from Equation (11), i.e.

$$\begin{aligned} G &= \sum_{j=1}^{N_B} \nu_j \left( \frac{S_{maxj} \, S_{aj}}{S_{a,max}^2} \right)^{k/2} \quad &\text{(a)} \\ G &= \sum_{j=1}^{N_B} \nu_j \left( \frac{S_{maxj}}{S_{a,max}} \sqrt{\frac{1-R_j}{2}} \right)^k \quad &\text{(b)} \\ G &= \sum_{j=1}^{N_B} \nu_j \left( \frac{S_{aj}}{S_{a,max}} \sqrt{\frac{2}{1-R_j}} \right)^k \quad &\text{(c)} \end{aligned} \tag{38}$$

Equation (38) returns equal to Equation (11) for R=−1 (i.e. if $S_{aj}=S_{maxj}$).

*3.2.2 Walker mean stress effect correction*

As done with the SWT model, if Walker equation (34) is substituted into (35) and the usual passages are performed, the following definition of G holds

$$G = \sum_{j=1}^{N_B} v_j \left(\frac{S_{maxj}^{1-\gamma} S_{aj}^{\gamma}}{S_{a,max}^2}\right)^k \quad (a)$$

$$G = \sum_{j=1}^{N_B} v_j \left(\frac{S_{maxj}}{S_{a,max}} \left(\frac{1-R_j}{2}\right)^{\gamma}\right)^k \quad (b) \qquad (39)$$

$$G = \sum_{j=1}^{N_B} v_j \left(\frac{S_{aj}}{S_{a,max}} \left(\frac{2}{1-R}\right)^{1-\gamma}\right)^k \quad (c)$$

Again, the first definition of G is retrieved under fully reversed loading. As regards the mean stress effect correction in the crack like notch and the blunt notch region, as well as the mean stress effect with a generic life dependent stress concentration factor $\varkappa(N)$, it is trivial to demonstrate that the current definition of G applies to Equation (21), (21) and (25), too.

## 4 Quantitative validation: SAE Keyhole test program

In the 1970s, the Society of Automotive Engineers (SAE) Fatigue Design & Evaluation Committee conducted a test program using a notched member with two steels commonly used in the ground vehicle industry (Bethlehem RQC-100 and U.S. Steel Man-Ten). The test program is explained in detail by Tucker and Bussa [63] and in the website https://www.efatigue.com/benchmarks/ under "SAE Keyhole Test Program", finally test program and analysis are described in the book [30]. Most basic material properties were measured for both materials (listed in Table 1). Constant amplitude tests were performed on the "component like" specimen, although the main scope of the test program was variable amplitude fatigue testing using three loading histories at several load levels. In fact, many different prediction models for constant and variable amplitude fatigue life have been collected in the SAE Transactions Vol. 84, 1975, § 1. For example, Landgraf et al. [64] and Potter [65] adopted a strain-life approach through Neuber's rule [66] whilst Nelson and Fuchs [67] decided to work with stress-life models called nominal stress range I and II methods. The TCD has been used lately also to predict with a certain level of accuracy the static failure of notched cold rolled low carbon steel and in presence of large plastic deformation before failure. Nonetheless, in order to state that the TCD is a fully established approach, a large number of tests is still needed on larger notch radii a> $a_0^u$ and on a wider amount of material classes; for this reason, albeit accurate in the cases analyzed, the current linear elastic approach does not expect to supersede the elastoplastic fracture mechanics. It is however noteworthy that no other simple methods amongst the ones collected in the SAE Transactions Vol. 84, 1975, § 1, including strain-life approach applied to notched geometries, seem to give higher accuracy.

Table 1 – Material properties for RQC-100 and Man-Ten

| Property Description | RQC-100 | Man Ten |
|---|---:|---:|
| Elastic Modulus, E, GPa | 203 | 203 |
| Yield Strength, Y, MPa | 883 | 325 |
| Ultimate Strength, $S_u$, MPA | 931 | 565 |
| Fatigue Limit Strength Range, $\Delta S_e$, MPa | 449 | 272 |
| Fatigue Strength Coefficient, $S'_f$, MPa | 1,240 | 1,160 |
| Threshold Stress Intensity Range, $\Delta K_{th}$, MPa·√mm | 158 | 285 |
| Fracture Toughness, $K_{Ic}$, MPa·√mm | 4,870 | 5,091 |
| Fatigue Strength Exponent, b | -0.07 | -0.095 |
| Fatigue Ductility Coefficient, ε | 1.06 | 0.26 |
| Fatigue Ductility Exponent, c | -0.75 | -0.47 |
| Crack Growth Intercept, C, mm·cycle$^{-1}$ | 5.2E-9 | 3.0E-9 |
| Crack Growth Exponent, m | 3.15 | 3.43 |

The specimen geometry and the test setup the SAE keyhole test program is provided in Figure 2, whilst the load set is given in Figure 3. From the load set adopted, a nominal stress $S_{nom}$ can be defined for convenience by "cutting" a beam shaped section immediately ahead of the crack tip and by writing the tensile stress from a combined axial-bending load (cfr. Figure 3). All the consideration related with the validity of the approximation used in the description of the stress field ahead of the notch root, stress concentration factors and toughness used in the setup of the TCD-P have been done and validated via comparison with finite elements analysis in Ciavarella et al. [33].

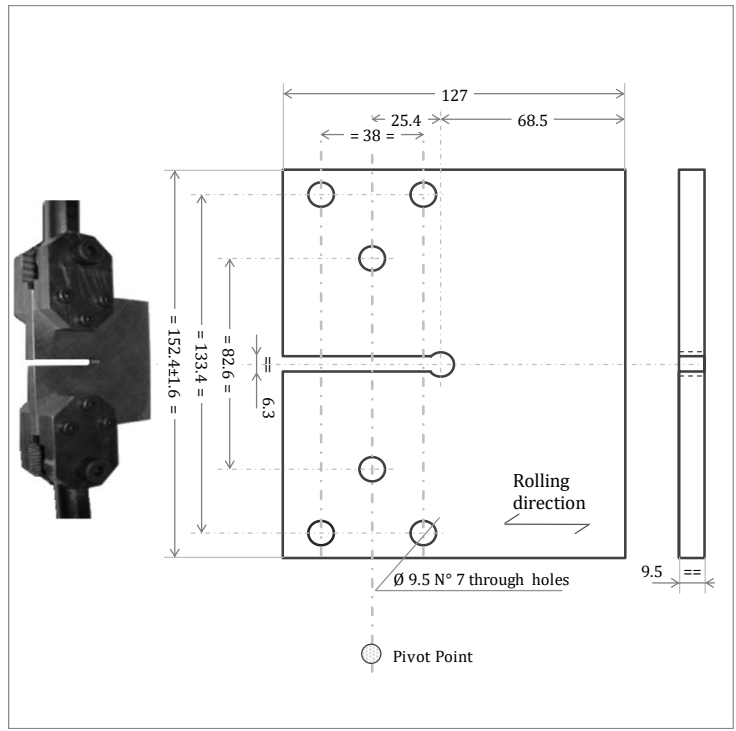

Figure 2 – SAE keyhole test specimen: (Left) experiment setup and (Right) dimensioned drawing. Units in mm

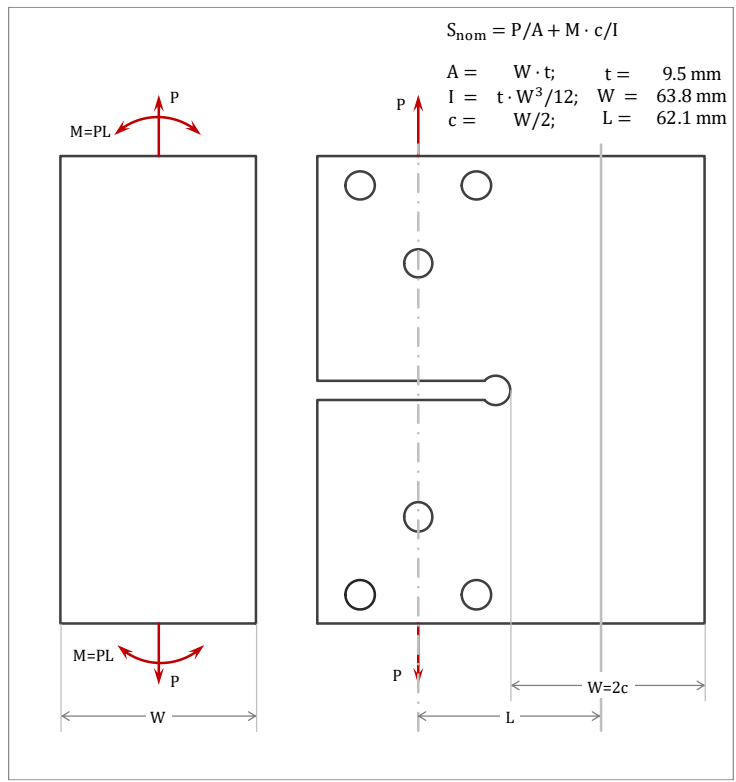

Figure 3 – SAE Keyhole test specimen: load set. The nominal stress has been calculated through the beam on the left

The CA life predictions are included here in Figure 4 for both RQC-100 (Left) and Man Ten (Right) to show the S/N curve models used in the analysis obtained through the TCD-P.

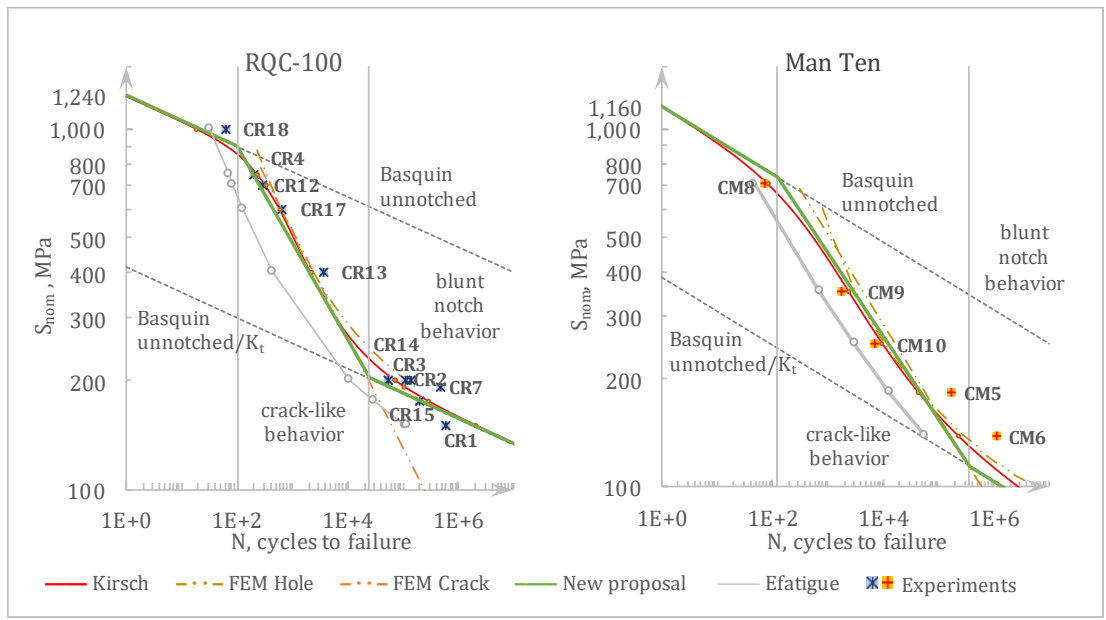

Figure 4 – SAE Keyhole test program: comparison of experimental data for constant amplitude fatigue of the RQC-100 (Left) and Man Ten (Right) specimen with the proposed TCD-P based model for notched or cracked specimen. (Left) Strain-life predictions are added to show. The alphanumeric code next to the experiment represents the type of spectrum, the material and the applied load

The loading histories used in the test program are (B) Bracket: narrow band load history, (T) Transmission: strong tensile bias with several compressive reversals and (S) Suspension: strong compressive bias. Some additional tests were done with the same truncated (mini) spectra, namely mB, mT, mS. All the spectra have been cycle counted through a rainflow algorithm (cfr. Matsuishi and Endo [68]) and shown here in Figure 5 with an unusual representation, i.e. not as a function of the time, but as contributions to the SWT mean stress corrected shift factor $G_j$. From Figure 5 it can be observed that (i) for higher non-dimensional stress amplitude there is usually a higher contribution to the $G_j$: indeed, G would have been a monotonic increasing function only if the loading spectra had been at R=−1; (ii) high tensile mean stresses tend to lower the shift factor $G^{-1/k}$ as expected; (iii) a higher Basquin's law exponent k tends to lower the $G^{-1/k}$. The VA life predictions have been calculated by shifting the green S/N curves shown in Figure 4 by $G^{-1/k}$ for each one of the six spectra and the results are plotted in Figure 6 for both RQC-100 (left) and for Man Ten (right). Predictions are satisfactory and are almost always in a scatter factor of ±3 times the predicted life. As regards RQC-100, almost all the predictions seem to be collocated on the conservative side of the bisector, while for Man Ten predictions are simply within the scatter bands.

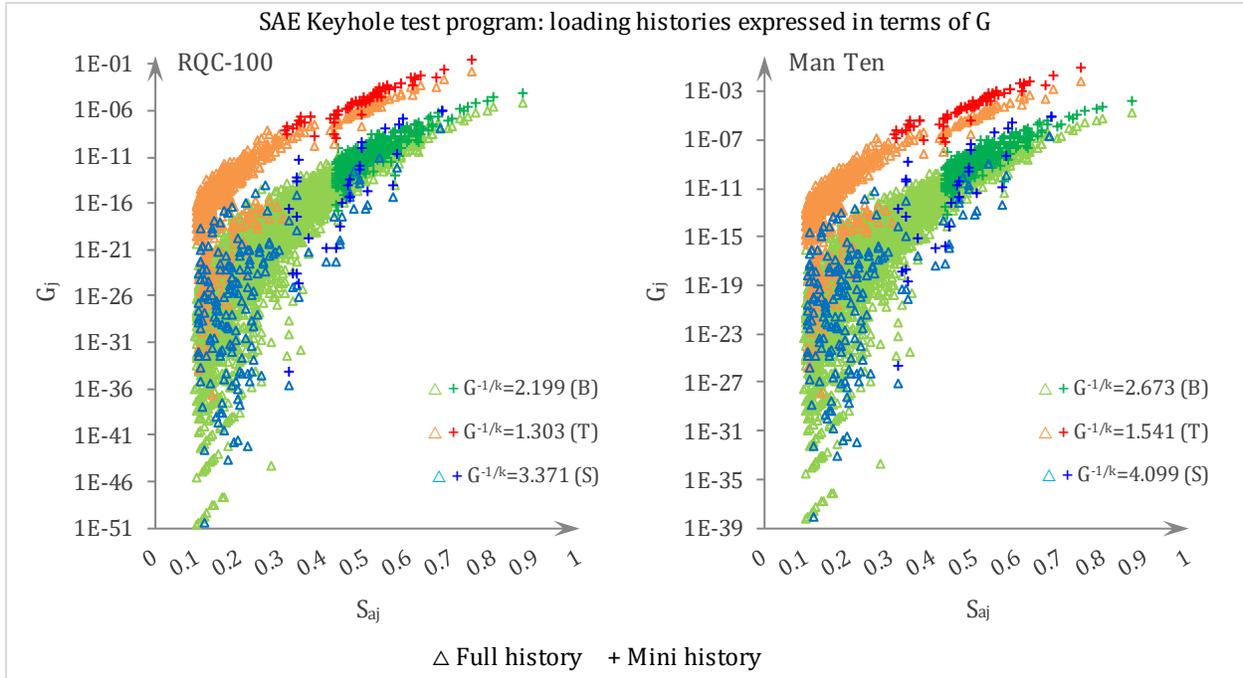

Figure 5 – SAE Keyhole test program: loading histories here expressed in terms of $G_j$, i.e. the j-th contribution to the shift factor for every material is here provided as a function of the non-dimensional stress amplitude. Letters B, T, S are the initials of the spectra: (B) Bracket, (T) Transmission (S) Suspension

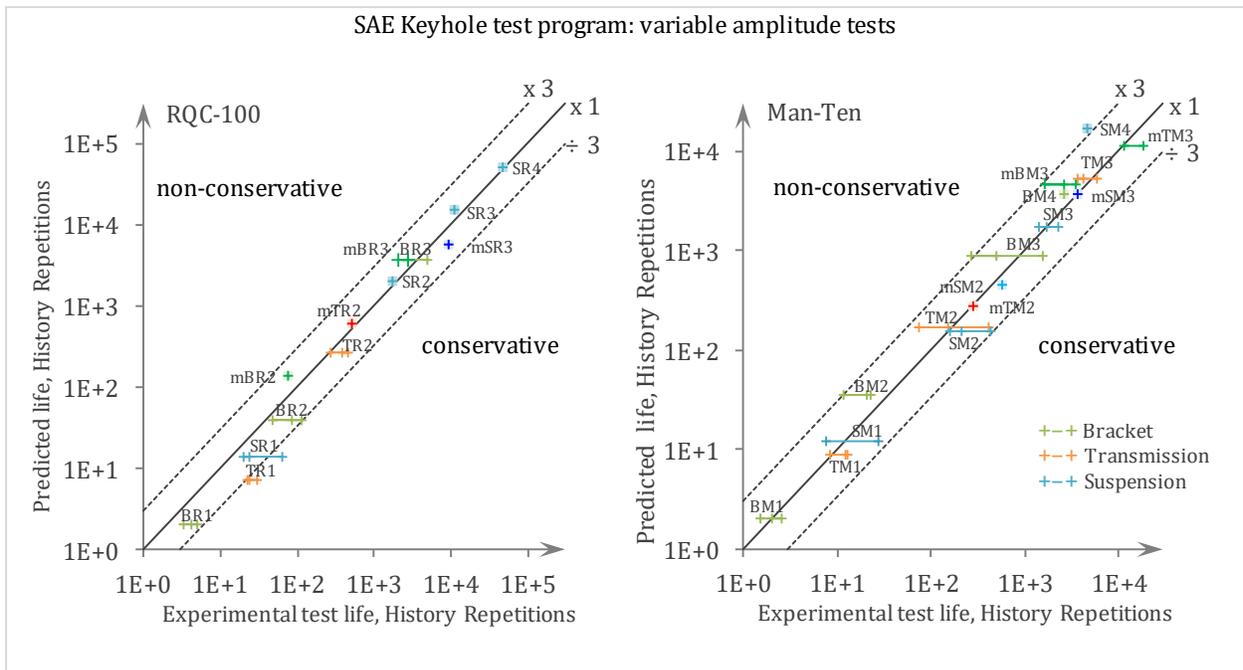

Figure 6 - SAE Keyhole test program: experimentally measured vs. predicted life through the shift factor approach. On the left predictions for RQC-100 and on the right for Man Ten. The solid line (bisector) is the prefect correspondence and the dashed lines are the scatter bands with multiplicative factors ±3. The alphanumeric codes are explained in Table 2

Table 2 – Specimen code number (from Tucker and Bussa [63])

| | | | |
|---|---|---|---|
| 1st letter | B | - | Bracket |
| History Identification | T | - | Transmission |
| | S | - | Suspension |
| | | | |
| 2nd letter | R | - | RQC-100 |
| Material Identification | M | - | Man Ten |
| | | | |
| 3rd number | 1 | - | Highest Load |
| | 2 | - | |
| | 3 | - | |
| | 4 | - | |
| | 5 | - | Lowest Load |

# 5 Discussion

## 5.1 Calibration of TCD-P constants

As already mentioned in the definition of Equation (5), there is a free parameter in the model: $N_u$, i.e. the number of cycles adopted as upper bound for $a_0(N)$, i.e. $a_0^u=a_0(N_u)$. In the cases under exam $N_u$ has been set to 1,000 cycles both for RQC-100 and for Man-Ten, which implies that $a_0^u=1/\pi\cdot(4{,}870/1{,}240)^2=4.91$ mm and $a_0^u=1/\pi\cdot(5{,}091/1{,}160)^2=6.13$ mm respectively. $N_u$ is the only fitting parameter the required by the model and it should be calibrated through best fitting technique. Other authors also have to recur to similar assumptions when calibrating the constants of the TCD method for finite life [19]. Notice that the plots here presented in Figure 4 have not been truncated below $N_u$ and above $N_e$ for cleanliness, albeit some truncations should be considered to have more precise plots. Thereupon, the TCD-P constants are:

$$B_{ST} = 2\frac{\text{Log}(158/4{,}870) + \text{Log}(1{,}240/449)}{\text{Log}(10^6/10^3)} \approx -0.70 \quad \text{(a)}$$

$$A_{ST} = \frac{1}{\pi}\left(\frac{158}{449}\right)^2 \cdot 10^{6\cdot 0.70} \approx 610 \text{ mm} \quad \text{(b)}$$

RQC-100 (40)

$$B_{ST} = 2\frac{\text{Log}(284/5{,}091) + \text{Log}(1{,}160/272)}{\text{Log}(10^6/10^3)} \approx -0.41 \quad \text{(a)}$$

$$A_{ST} = \frac{1}{\pi}\left(\frac{285}{272}\right)^2 \cdot 10^{6\cdot 0.42} \approx 100 \text{ mm} \quad \text{(b)}$$

Man Ten (41)

Notice that, using Ciavarella's proposal [27] and taking r equal to the Paris' law exponent, the constants would be comparable for Man Ten but quite different for RQC-100:

$$B_C = 2\left(\frac{1}{14.28} - \frac{1}{3.15}\right) \approx -0.50 \quad (a)$$

$$A_C = \frac{1}{\pi}\left(\frac{158}{449}\right)^2 \cdot 10^{6 \cdot 0.50} \approx 40 \text{ mm} \quad (b)$$

RQC-100 (42)

$$B_C = 2\left(\frac{1}{10.53} - \frac{1}{3.43}\right) \approx -0.40 \quad (a)$$

$$A_C = \frac{1}{\pi}\left(\frac{285}{272}\right)^2 \cdot 10^{6 \cdot -0.40} \approx 74 \text{ mm} \quad (b)$$

Man Ten (43)

In order to have $A_C$ and $B_C$ equal to $A_{ST}$ and $B_{ST}$, the exponent r should be equal to 2.4 for RQC-100 and 3.3 for Man Ten.

## 6   Conclusion

It has been demonstrated the S/N curves under CA or VA loading can be obtained by a simple shift factor depending on the spectrum histogram and the Basquin's law slope, within the assumptions of PM rule and TCD-P. This holds true for both smooth and notched specimens as demonstrated by the authors in a previous paper. The finding is based on a TCD simple method proposed and validated by Susmel and Taylor. However, considering this result, there is no need to apply the iterative calculations that Susmel and Taylor propose, as the VA curves can be obtained directly from the CA curves, for which many proposals have already been put forward, also in closed form. Even the computation of the stress field from Finite Element Method does not seem necessary in many cases, as it does not add much accuracy to a problem where the number of assumptions is already quite strong, and more important, than the details of the stress field. As a first approximation, spectrum loading effects in notched or even cracked structures can be estimated easily from reduced amount of testing. Mean stress effect correction has been introduced in the definition of the shift factor through the SWT and the Walker equations and it has been demonstrated that under fully reversed loading the initial definition of shift factor is retrieved. The findings have finally been applied to the SAE Keyhole test program data, from which VA fatigue life experiments for two materials under three different complex loading spectra were available. In all the cases predictions have provided satisfactory accuracy.

# References

[1] S. Suresh, *Fatigue of Materials*. Cambridge University Press, 1998.
[2] R. C. Juvinall and K. M. Marshek, *Fundamentals of Machine Component Design*. Wiley, 2017.
[3] R. I. Stephens, A. Fatemi, R. R. Stephens, and H. O. Fuchs, *Metal Fatigue in Engineering*. John Wiley & Sons, 2000.
[4] H. O. Fuchs and R. I. Stephens, *Metal fatigue in engineering*. Wiley, 1980.
[5] D. Radaj, *Ermüdungsfestigkeit (Fatigue strength, in German), 2003*. Springer, Berlin, Heidelberg, New York.
[6] P. P. Milella, *Fatigue and Corrosion in Metals*. Mailand: Springer-Verlag, 2013.
[7] M. M. Pedersen, 'Introduction to Metal Fatigue', *Technical Report Mechanical Engineering*, vol. 5, no. 11, pp. 91–91, Nov. 2018.
[8] J. H. S. Redfern, 'An approach to metal fatigue', Jun. 1965.
[9] W. Weibull, *Fatigue Testing and Analysis of Results*. Elsevier, 1961.
[10] H. P. Rossmanith, 'George Rankin Irwin-The Father of Fracture Mechanics 1907-1998', *Fragblast*, vol. 2, no. 2, pp. 123–141, Jan. 1998.
[11] R. A. Smith and K. J. Miller, 'Prediction of fatigue regimes in notched components', *International Journal of Mechanical Sciences*, vol. 20, no. 4, pp. 201–206, 1978.
[12] B. Atzori and P. Lazzarin, 'Analisi delle problematiche connesse con la valutazione numerica della resistenza a fatica', in *AIAS National Conference, Lucca Italy, also Quaderno AIAS*, 2000, pp. 33–50.
[13] B. Atzori and P. Lazzarin, 'Notch Sensitivity and Defect Sensitivity under Fatigue Loading: Two Sides of the Same Medal', *International Journal of Fracture*, vol. 107, no. 1, pp. 1–8, Jan. 2001.
[14] B. Atzori and P. Lazzarin, 'A three-dimensional graphical aid to analyze fatigue crack nucleation and propagation phases under fatigue limit conditions', *International Journal of Fracture*, vol. 118, no. 3, pp. 271–284, Dec. 2002.
[15] B. Atzori, P. Lazzarin, and G. Meneghetti, 'Fracture mechanics and notch sensitivity', *Fatigue & Fracture of Engineering Materials & Structures*, vol. 26, no. 3, pp. 257–267, 2003.
[16] B. Atzori, G. Meneghetti, and P. Lazzarin, 'Fatigue and Notch Mechanics', *Fatigue and Notch Mechanics*, 17-Jul-2019.
[17] P. C. Paris and F. Erdogan, *A Critical Analysis of Crack Propagation Laws*. ASME, 1963.
[18] T. Nicholas, *High Cycle Fatigue: A Mechanics of Materials Perspective*. Elsevier Science, 2006.
[19] H. Kitagawa and S. Takahashi, 'Applicability of fracture mechanics to very small cracks or the cracks in the early stage', in *Proc. of 2nd ICM, Cleveland, 1976*, 1976, pp. 627–631.
[20] M. H. El Haddad, N. E. Dowling, T. H. Topper, and K. N. Smith, 'J integral applications for short fatigue cracks at notches', *International Journal of Fracture*, vol. 16, no. 1, pp. 15–30, 1980.
[21] D. Taylor, 'The theory of critical distances', *Engineering Fracture Mechanics*, vol. 75, no. 7, pp. 1696–1705, 2008.
[22] W. Yao, K. Xia, and Y. Gu, 'On the fatigue notch factor, Kf', *International Journal of Fatigue*, vol. 17, no. 4, pp. 245–251, May 1995.
[23] H. Neuber, *Theory of notch stresses: Principles for exact stress calculation*, vol. 74. JW Edwards, 1946.
[24] P. Kuhn and H. F. Hardrath, 'An engineering method for estimating notch-size effect in fatigue tests on steel', National Advisory Committee for Aeronautics, Langley Field, Va, NACA Technical Note NACA-TR-2805, 1952.
[25] N. Pugno, M. Ciavarella, P. Cornetti, and A. Carpinteri, 'A generalized Paris' law for fatigue crack growth', *Journal of the Mechanics and Physics of Solids*, vol. 54, no. 7, pp. 1333–1349, Jul. 2006.
[26] M. Ciavarella and F. Monno, 'On the possible generalizations of the Kitagawa–Takahashi diagram and of the El Haddad equation to finite life', *International Journal of Fatigue*, vol. 28, no. 12, pp. 1826–1837, Dec. 2006.
[27] M. Ciavarella, 'Crack propagation laws corresponding to a generalized El Haddad equation', *International Journal of Aerospace and Lightweight Structures (IJALS)*, vol. 1, no. 1, 2011.
[28] J. Maierhofer, H.-P. Gänser, and R. Pippan, 'Modified Kitagawa–Takahashi diagram accounting for finite notch depths', *International Journal of Fatigue*, vol. 70, pp. 503–509, Jan. 2015.

[29] L. Susmel and D. Taylor, 'A novel formulation of the theory of critical distances to estimate lifetime of notched components in the medium-cycle fatigue regime', *Fatigue & Fracture of Engineering Materials & Structures*, vol. 30, no. 7, pp. 567–581, 2007.

[30] L. Susmel and D. Taylor, 'On the use of the Theory of Critical Distances to predict static failures in ductile metallic materials containing different geometrical features', *Engineering Fracture Mechanics*, vol. 75, no. 15, pp. 4410–4421, Oct. 2008.

[31] L. Susmel and D. Taylor, 'The Theory of Critical Distances to estimate lifetime of notched components subjected to variable amplitude uniaxial fatigue loading', *International Journal of Fatigue*, vol. 33, no. 7, pp. 900–911, Jul. 2011.

[32] L. Susmel and D. Taylor, 'A critical distance/plane method to estimate finite life of notched components under variable amplitude uniaxial/multiaxial fatigue loading', *International Journal of Fatigue*, vol. 38, pp. 7–24, May 2012.

[33] M. Ciavarella, P. D'Antuono, and G. P. Demelio, 'Generalized definition of "crack-like" notches to finite life and SN curve transition from "crack-like" to "blunt notch" behavior', *Engineering Fracture Mechanics*, vol. 179, pp. 154–164, Jun. 2017.

[34] A. G. Palmgren, 'Die Lebensdauer von Kugellagern (Life Length of Roller Bearings. In German)', *Zeitschrift des Vereines Deutscher Ingenieure (VDI Zeitschrift), ISSN*, pp. 0341–7258, 1924.

[35] B. F. Langer, 'Fatigue failure from stress cycles of varying amplitude', *Journal of Applied Mechanics*, vol. 59, pp. A160–A162, 1937.

[36] M. A. Miner, 'Cumulative Damage in Fatigue', . *Journal of Applied Mechanics*, vol. 3, pp. 159–164, 1945.

[37] M. Ciavarella, P. D'antuono, and A. Papangelo, 'On the connection between Palmgren-Miner rule and crack propagation laws', *Fatigue & Fracture of Engineering Materials & Structures*, vol. 41, no. 7, pp. 1469–1475, 2018.

[38] H. L. Leve, 'Cumulative damage theories', *Metal Fatigue: Theory and Design*, vol. A. F. Madayag Ed., pp. 170–203, 1960.

[39] C. Park and W. J. Padgett, 'A general class of cumulative damage models for materials failure', *Journal of Statistical Planning and Inference*, vol. 136, no. 11, pp. 3783–3801, Nov. 2006.

[40] E. Haibach, *Analytical Strength Assessment of Components in Mechanical Engineering: FKM-Guideline*. VDMA, 2003.

[41] C. M. Sonsino, 'Principles of variable amplitude fatigue design and testing', in *Fatigue Testing and Analysis Under Variable Amplitude Loading Conditions*, ASTM International, 2005.

[42] C. M. Sonsino and K. Dieterich, 'Fatigue design with cast magnesium alloys under constant and variable amplitude loading', *International Journal of Fatigue*, vol. 28, no. 3, pp. 183–193, Mar. 2006.

[43] C. M. Sonsino, 'Fatigue testing under variable amplitude loading', *International Journal of Fatigue*, vol. 29, no. 6, pp. 1080–1089, Jun. 2007.

[44] J. Schijve, *Fatigue of Structures and Materials*. Springer Science & Business Media, 2001.

[45] J. Schijve, Ed., 'Fatigue under Variable-Amplitude Loading', in *Fatigue of Structures and Materials*, Dordrecht: Springer Netherlands, 2009, pp. 295–328.

[46] J. Tomblin and W. Seneviratne, 'Determining the fatigue life of composite aircraft structures using life and load-enhancement factors. Final report', 2011.

[47] E. Haibach, 'Betriebsfestigkeit: Verfahren und Daten zur Bauteilberechnung. Düsseldorf: VDI-Verlag', ISBN 3–18–400828–2, 1989.

[48] M. Ciavarella, P. D'Antuono, and G. P. Demelio, 'A simple finding on variable amplitude (Gassner) fatigue SN curves obtained using Miner's rule for unnotched or notched specimen', *Engineering Fracture Mechanics*, vol. 176, pp. 178–185, May 2017.

[49] H. M. Westergaard, 'Stresses At A Crack, Size Of The Crack, And The Bending Of Reinforced Concrete', *JP*, vol. 30, no. 11, pp. 93–102, Nov. 1933.

[50] M. Ciavarella, 'A simple approximate expression for finite life fatigue behaviour in the presence of "crack-like" or "blunt" notches', *Fatigue & Fracture of Engineering Materials & Structures*, vol. 35, no. 3, pp. 247–256, 2012.

[51] A. Wöhler, 'Bericht über die Versuche, welche auf der Königl. Niederschlesisch-Märkischen Eisenbahn mit Apparaten zum Messen der Biegung und Verdrehung von Eisenbahnwagen-Achsen während der Fahrt angestellt wurden', *Zeitschrift für Bauwesen*, vol. 8, no. 1858, pp. 641–652, 1858.

[52] A. Wöhler, 'Versuche zur Ermittlung der auf die Eisenbahnwagenachsen einwirkenden Kräfte und die Widerstandsfähigkeit der Wagen-Achsen', *Zeitschrift für Bauwesen*, vol. 10, no. 1860, pp. 583–614, 1860.

[53] A. Wöhler, *Ueber die Festigkeits-versuche mit Eisen und Stahl*. 1870.

[54] H. Gerber, *Bestimmung der zulässigen Spannungen in Eisen-Constructionen*. Wolf, 1874.

[55] J. Goodman, *Mechanics applied to engineering*. London [etc.] Longmans, Green & Co., 1914.

[56] C. R. Söderberg, 'Factor of safety and working stress', *Trans Am Soc Mech Eng*, vol. 52, pp. 13–28, 1939.

[57] A. R. Woodward, K. W. Gunn, and G. Forrest, 'The effect of mean stress on the fatigue of aluminum alloys', in *International Conference on Fatigue of Metals*, 1956, pp. 1156–1158.

[58] N. E. Dowling, 'Mean Stress Effects in Stress-Life and Strain-Life Fatigue', SAE International, Warrendale, PA, SAE Technical Paper 2004-01–2227, Apr. 2004.

[59] N. E. Dowling, C. A. Calhoun, and A. Arcari, 'Mean stress effects in stress-life fatigue and the Walker equation', *Fatigue fract. eng. mater. struct. (Print)*, vol. 32, no. 3, pp. 163–179, 2009.

[60] J. Morrow, 'Fatigue properties of metals', *Fatigue design handbook*, pp. 21–30, 1968.

[61] K. N. Smith, P. Watson, and T. H. Topper, 'A Stress-Strain Function for the Fatigue of Metals', *Journal of Materials*, vol. 5, pp. 767–778, Dec. 1970.

[62] K. Walker, 'The Effect of Stress Ratio During Crack Propagation and Fatigue for 2024-T3 and 7075-T6 Aluminum', *Effects of Environment and Complex Load History on Fatigue Life*, Jan. 1970.

[63] L. Tucker and S. Bussa, 'The SAE Cumulative Fatigue Damage Test Program', *SAE Transactions*, vol. 84, pp. 198–248, 1975.

[64] R. W. Landgraf, F. D. Richards, and N. R. LaPointe, 'Fatigue Life Predictions for a Notched Member Under Complex Load Histories', *SAE Transactions*, vol. 84, pp. 249–259, 1975.

[65] J. M. Potter, 'Spectrum Fatigue Life Predictions for Typical Automotive Load Histories and Materials Using the Sequence Accountable Fatigue Analysis', *SAE Transactions*, vol. 84, pp. 260–269, 1975.

[66] H. Neuber, 'Theory of stress concentration for shear-strained prismatical bodies with arbitrary nonlinear stress-strain law', *Journal of applied mechanics*, vol. 28, no. 4, pp. 544–550, 1961.

[67] D. V. Nelson and H. O. Fuchs, 'Predictions of Cumulative Fatigue Damage Using Condensed Load Histories', *SAE Transactions*, vol. 84, pp. 276–299, 1975.

[68] M. Matsuishi and T. Endo, 'Fatigue of metals subjected to varying stress', *Japan Society of Mechanical Engineers, Fukuoka, Japan*, vol. 68, no. 2, pp. 37–40, 1968.